\documentclass{article}
\usepackage[paper=letterpaper,margin=2cm]{geometry}
\usepackage{soul, color, xcolor}
\usepackage{graphicx}
\usepackage{amsmath}
\usepackage{amssymb}
\usepackage{braket}
\usepackage{bm}
\usepackage{amsfonts}
\usepackage{cite}
\usepackage[ruled,boxed,linesnumbered]{algorithm2e}
\usepackage{tcolorbox}
\usepackage[colorlinks=true,citecolor=blue]{hyperref}
\usepackage{booktabs}
\usepackage{extarrows}
\tcbuselibrary{breakable}
\date{}
\title{\textbf{Entanglement Model for Mode-Pairing \\Quantum Key Distribution}}

\author{\begin{minipage}{0.92\textwidth}
    \centering
    \small
    Yi-Fei Lu$^{1,2}$, Yang Wang$^{1,2,3,\dagger}$, Hong-Wei Li$^{1,2}$, Mu-Sheng Jiang$^{1,2}$, Xiao-Xu Zhang$^{1,2}$, Ying-Ying Zhang$^{1,2,4}$, Yu Zhou$^{1,2}$, Xiao-Lei Jiang$^{1,2}$, Hai-Tao Wang$^{1,2}$, Yan-Mei Zhao$^{1,2}$, Chun Zhou$^{1,2}$, Wan-Su Bao$^{1,2,\ddagger}$\\
    $^1$Henan Key Laboratory of Quantum Information and Cryptography, IEU, Zhengzhou 450001, China\\
    $^2$Synergetic Innovation Center of Quantum Information and Quantum Physics, \\University of Science and Technology of China, Hefei 230026, China\\
    $^3$National Laboratory of Solid State Microstructures, School of Physics and Collaborative Innovation Center of Advanced Microstructures, Nanjing University, Nanjing 210093, China\\
    $^4$Xi'an Satellite Control Center, Xian 710043, China\\
    $^\dagger$wy@qiclab.cn,  $^\ddagger$bws@qiclab.cn
\end{minipage}}

\begin{document}
\maketitle

{\centering\section*{Abstract}}

Mode-pairing (MP) quantum key distribution (QKD) eliminates the requirements of phase locking and phase tracking compared with twin-field (TF) QKD while still surpassing the fundamental rate-distance limit of QKD. The complexity of the experimental implementation is reduced while the efficiency can still be guaranteed. In MP-QKD, two communication parties need to pair two effective rounds according to the announced results by the third party. Therefore, it is not intuitive how to provide an entanglement protocol equivalent to the prepare-and-measure protocol for proving security. At present, the security of MP-QKD is rigorously proven by examining the consistency of the states detailly between MP-QKD and the fixed-pairing scheme under all of Eve's possible interference to obtain the equivalence, and verifying the security of the latter. 
Here, we directly present an entanglement model for MP-QKD by proposing a free-pairing entanglement scheme that is equivalent to MP-QKD. This entanglement model simplifies the security proof and provides a clearer insight into the foundation of MP-QKD. Besides, it could provide a theoretical framework for the comprehensive analysis of MP-QKD, such as facilitating practical security analysis. Additionally, we propose an optimized pairing strategy based on the entanglement model. The simulation results show an enhancement of 163\% in pairing efficiency with a reasonable pairing interval $2 \times 10^5$.

\clearpage

%------------------------------
%set the box
\tcbset{colbacktitle=yellow!10!white, colback=white,coltitle=black, breakable, before upper={\parindent10pt\noindent}, fonttitle=\bfseries}
%------------------------------

\section{Introduction}

Quantum key distribution (QKD) provides a method for distributing secret key bits with information-theoretically security guaranteed by the laws of quantum physics \cite{bennett2014RN153,lo1999RN138,shor2000RN300}. However, rigorous security requires certain implementation assumptions that may not be met in practical systems. Furthermore, the practical performance, including secret key rate and distance, is limited by the channel loss. Many theoretical and experimental breakthroughs have been made to overcome these practical challenges \cite{xu2020RN130, portmann2022RN813}.

The decoy-state method \cite{hwang2003RN202,lo2005RN81,wang2005RN185} enables practical QKD with weak coherent pulses by characterizing the quantum channel with additional states. This approach can overcome photon-number-splitting attacks \cite{brassard2000RN299,lutkenhaus2002RN301} and achieve a secret key rate comparable to that of the single-photon source. The measurement-device-independent (MDI) QKD \cite{lo2012RN72,braunstein2012RN581} enhances the practical security of QKD by eliminating all potential security vulnerabilities at the detection side. In MDI-QKD, Alice and Bob are both located at the source side, while an untrusted party named Charlie performs the measurement procedure. Charlie can only infer the parity of Alice and Bob's bits but not their specific values. From the perspective of entanglement purification, Alice and Bob's relationship is established through entanglement swapping with Charlie's Bell state measurement. The combination of the decoy-state method and MDI-QKD significantly enhances both practicality and practical security~\cite{liu2013RN900,yin2016RN288}.

The practical performance of QKD is limited by the optical loss in the quantum channel. There is an upper bound of the secret key rate, which exponentially decreases as the distance increases. For example, the PLOB bound \cite{pirandola2017RN103} characterizes the fundamental rate-distance limit of QKD without quantum repeaters. However, the proposal of twin-field (TF) QKD \cite{lucamarini2018RN45,ma2018RN56,wang2018RN22,lin2018RN391,tamaki2018RN280,curty2019RN57,cui2019RN41,maeda2019RN507} has successfully overcome this limit based on the single-photon interference. The core of TF-QKD is entanglement swapping, which is similar to MDI-QKD but uses coherent states instead of single-photon states as carriers. Many experimental breakthroughs have been achieved currently~\cite{minder2019RN46,zhong2019RN52,liu2019RN19,fang2020RN55,liu2021RN490,chen2021RN586,pittaluga2021RN482,chen2022RN735,wang2022RN585,clivati2022RN579,chang2022RN738, li2023RN979, zhou2023RN949}. However, TF-QKD encodes the key information into the phase of coherent states and requires technologies for phase tracking and phase locking to align the phase reference frame and compensate for the phase drift, which is a complex and time-consuming process.

To remove these technologies from TF-QKD, mode-pairing (MP) QKD (also known as asynchronous MDI-QKD) was simultaneously proposed in Refs. \cite{zeng2022RN816} and \cite{xie2022RN724}, which can surpass the fundamental rate-distance limit of QKD without phase tracking and phase locking. MP-QKD is an enhanced version of the time-bin encoding MDI-QKD~\cite{ma2012RN73} by eliminating the requirement that coincidence detection events are only sifted between neighboring rounds, i.e., the paired rounds are decoupled. Therefore, the secret key rate is promoted as numerous effective rounds without neighbors can be recycled, resulting in the scaling of $O(\sqrt{\eta})$ when the maximal pairing interval $l \to \infty$ \cite{zeng2022RN816}. This means that MP-QKD can enjoy both practicality and efficiency. To demonstrate the power of MP-QKD, many experiments have been successfully performed in the laboratory \cite{zhu2023RN928,zhou2023RN978,zhu2024RN1180,ge2025RN1181}, obtaining a quadratic improvement in the key rate. At present, many analyses of MP-QKD with practical issues have been presented to promote its practicality \cite{wang2023RN1051,bai2023RN1184,liu2023RN1182,xie2023RN1185,xie2023RN1097,lu2024RN1183,zhou2024RN1084,2024RN1176}.

In MP-QKD, Alice and Bob need to pair two effective rounds (e.g. $j$ and $k$) as a pair to encode bits, which are determined by the third party, Charlie. Note that MP-QKD is transformed into time-bin encoding MDI-QKD when the round $k$ is fixed as $j+1$. The decoupling of the pairing rounds $j$ and $k$ is a major difference between MP-QKD and MDI-QKD, which is essential for the security of MP-QKD since the rounds $j$ and $k$ are determined by the third untrusted party in MP-QKD. The usual approach to guaranteeing the security of QKD is to provide an entanglement protocol that is equivalent to the prepare-and-measure protocol and prove the security of the former. When the pairing strategy is fixed, meaning it is not affected by measurement results and is determined locally by Alice and Bob, the entanglement protocol can be obtained by setting a proper entangled state between the paired rounds of Alice and Bob's systems in the state preparation step. However, the pairs in MP-QKD are determined by the announced measurement results, and the pairing results for Alice and Bob are random. Therefore, it is not intuitive to provide an entanglement protocol. Ref. \cite{zeng2022RN816} provides a detailed security proof by first proving the security of fixed-pairing entanglement scheme, where the entangled states are assumed in every round. Its reduction to the prepare-and-measure scheme is directly obtained since Alice and Bob's operations are independent of Charlie's announced detection results under the fixed pairing condition. Then it proves the equivalence of the fixed-pairing and free-pairing schemes by noting that there are no difference of Alice and Bob's states and Charlie's possible operations. In this way, it proves the security of MP-QKD.

In this work, we aim to directly provide an free-pairing entanglement scheme that is equivalent to the prepare-and-measure scheme for MP-QKD. We assume the entangled states between the ancillary and actual systems for every round in the state-preparation step. Specifically, the state $\ket{0}$ corresponds to the vacuum state and $\ket{1}$ corresponds to a coherent pulse. To establish the entanglement between the paired rounds based on Charlie's announced results, we introduce a positive operator-valued measure (POVM) on the paired auxiliary systems after Charlie's measurement. With the POVM, we can clearly understand how entanglement is established in paired rounds based on Charlie’s announced results. We prove the security of this free-pairing entanglement scheme by analyzing the distillable entanglement based on the tagging method. To reduce it to the prepare-and-measure scheme, we prove the commutativity of Alice and Bob's measurements with Charlie's measurement. The main challenge to prove the commutativity lies in the fact that Alice and Bob’s operations are based on Charlie’s announced results. To address this issue, we combine and simplify Alice and Bob’s operations after Charlie's announcement to a simple measurement, which is commute with Charlie's measurement. In this way, we prove the equivalence of the entanglement scheme and prepare-and-measure scheme and complete the security proof. With this equivalence, it is beneficial to understand why the rounds $j$ and $k$ can be decoupled and how entanglement between Alice and Bob is established. The entanglement model provides a theoretical framework for the analysis of MP-QKD. As an application, we analyze and improve the pairing strategy based on the entanglement model. The simulation results show an enhancement of 163\% in pairing efficiency with a reasonable pairing interval $2 \times 10^5$ \cite{zhou2023RN978}, which improves the secret key rate. In addition, this model could serve as an essential tool for practical security analysis.

In Sec. \ref{protocol}, we introduce the MP-QKD protocol and provide some notations. We propose an free-pairing entanglement scheme for MP-QKD, prove its security, and reduce it to the prepare-and-measure scheme in Sec. \ref{proof}. In Sec. \ref{analysis}, we analyze and propose an improved pairing strategy for decoy-state MP-QKD. Finally, the conclusion is given in Sec. \ref{conclusion}.

\section{MP-QKD Protocol}
\label{protocol}

In time-bin encoding MDI-QKD, Alice and Bob prepare the paired states at the beginning, which means that the states in two paired rounds are correlated in the state preparation stage. However, in MP-QKD, the paired rounds are determined by Eve's measurement results, and the states in different rounds are prepared independently because Alice and Bob can not predict the measurement results during the state preparation stage. Nevertheless, Alice and Bob need to correlate them according to the measurement results, resulting in differences between MDI-QKD and MP-QKD. In this section, we present the MP-QKD protocol and the pairing strategy \cite{zeng2022RN816,xie2022RN724} and provide some notations.

\begin{tcolorbox}[title = {Box 1: MP-QKD \cite{zeng2022RN816, xie2022RN724}}]

    (1) State preparation. In $k$-th round, Alice chooses a random bit $a_k \in\mathbb{Z}_2$ and a random phase $\theta^a_k \in [0,2\pi)$. Then Alice prepare a vacuum state $\ket{0}$ or coherent state $\ket{e^{i\theta^a_k}\sqrt{\mu}}$ when $a_k = 0$ or $1$. Similarly, Bob independently chooses bit $b_k \in\mathbb{Z}_2$ and phase $\theta^b_k \in [0,2\pi)$ and performs the same procedure. Here we omit the details of the decoy-state method for simplicity.

    (2) Measurement. Alice and Bob send the prepared states to the third party, Charlie, who is supposed to perform the interferometric measurements and announce the results $L_k, R_k \in\mathbb{Z}_2$ of two detectors. The results 1 and 0 denote whether Charlie announced a detector click or not.

    (3) Mode pairing. After repeating the above steps $N$ times, Alice and Bob sift the effective rounds when $L_k \oplus R_k=1$ and perform the pairing according to the pairing strategy in Box 2.

    (4) Basis sifting. Alice and Bob independently assign the pairs (indexed by $j$ and $k$) as $Z$ basis if the intensities are $(0,\mu)$ or $(\mu,0)$, as $X$ basis if $(\mu,\mu)$, or as '0' if $(0,0)$. They announce the basis and sift the only pairs when both are $Z$ to perform the key mapping.

    (5) Key mapping. Alice and Bob set their raw key bits as $\alpha_{jk}=a_j \overline{a}_k$ and $\beta_{jk}=\overline{b}_j b_k$ in $Z$ basis, respectively. Alice and Bob announce their phase difference $\delta_{ij}^a = \theta_{j}^a - \theta_{k}^a$ and $\delta_{ij}^b = \theta_{j}^b - \theta_{k}^b$ in $X$ basis, which can be used to estimate the phase error rate.

    (6) Parameter estimation. By employing the decoy-state method, they estimate the fraction $q_{11}^z$ of single-photon pairs when both Alice and Bob send the single-photon states $\ket{01}$ or $\ket{10}$ among $Z$ basis pairs when both Alice and Bob are $Z$ basis, and phase error rate $e^x_{11}$ of the single-photon pairs.

    (7) Key distillation. Alice and Bob perform error correction and privacy amplification to distill the final key bits. The secret key rate is given by \cite{gottesman2004RN266,zeng2022RN816}
    \begin{equation}
        \label{rate}
        R = r_z \big\{ q_{11}^z [1 - h(e^x_{11})] - \lambda_{\text{EC}} \big\},
    \end{equation}
    where $r_z$ is the pairing efficiency of $Z$ basis among all rounds $N$, $h(x)=-x\log_2x -(1-x)\log_2(1-x)$ is the binary Shannon entropy function, $\lambda_{\text{EC}}$ is the information revealed in the error correction step. Here, $r_z$ and $\lambda_{\text{EC}}$ can be obtained directly in practical application.

\end{tcolorbox}

The pairing strategy is shown in Box 2 \cite{zeng2022RN816}. The main idea is to pair neighboring effective rounds within the maximal pairing interval. The maximal pairing interval $l$ represents the maximum distance between two paired rounds that satisfy condition $k-j\leq l$. This interval is determined by the phase drift rate, which we consider as a parameter without analyzing how to determine $l$ based on experiments. Detailed experimental analysis can be found in Refs. \cite{zeng2022RN816,xie2022RN724}.

\begin{tcolorbox}[title = {Box 2: Pairing Strategy \cite{zeng2022RN816}}]

    \textbf{Input}: Charlie's announced detection results $C_i = L_i \oplus R_i$ for $i=1$ to $N$; maximal pairing interval $l$.

    \textbf{Output}: $K$ pairs, $(F_k,R_k)$ for the $k$-th pair for $k=1$ to $K$.

    \textbf{Initialization}: $k=1$; $f=0$.

    \textbf{for} $i\in[N]$ \textbf{do}

    \quad \textbf{if} $C_i=1$ \textbf{then}

    \quad\quad \textbf{if} $f=0$ \textbf{then}

    \quad\quad\quad $F_k\leftarrow i$; $f\leftarrow 1$.

    \quad\quad \textbf{else then}

    \quad\quad\quad\ \textbf{if} $i-F_k\leq l$ \textbf{then}

    \quad\quad\quad\quad $R_k\leftarrow i$; $k\leftarrow k+1$, $f\leftarrow 0$.

    \quad\quad\quad\ \textbf{else then} 
    
    \quad\quad\quad\quad $F_k\leftarrow i$.

    \quad\quad\quad \textbf{end if}

    \quad\quad \textbf{end if}

    \quad \textbf{end if}

    \textbf{end for}
    
\end{tcolorbox}

\section{Security Proof of MP-QKD}
\label{proof}

To prove the security of MP-QKD, it is crucial to explain how Alice and Bob can correlate the states in paired rounds that are prepared independently, based on Eve's measurement results. In this section, we first present an free-pairing entanglement scheme for MP-QKD, then prove its security, and finally reduce it to the prepare-and-measure scheme for MP-QKD. We can directly analyze the entanglement protocol to investigate the performance and security of MP-QKD.

\subsection{Entanglement Scheme for MP-QKD}
\label{entanglement}

To provide an entanglement scheme for MP-QKD, we introduce ancillary systems that are locally kept at Alice and Bob to create extended states. We take Alice as an example and the analysis for Bob is similar.

In step-1, Alice determines a random bit as $0$ or $1$ and prepares the vacuum state $\ket{0}$ or the coherent state $\ket{e^{i\theta^a_k}\sqrt{\mu}}$, respectively. The prepared states can be expressed as
\begin{equation}
    \rho_{A_k} = \frac{1}{2} \big[\hat{P}\big(\ket{0}_{A_k}\big) + \hat{P}\big(\ket{e^{i\theta^a_k}\sqrt{\mu}}_{A_k}\big)\big],
\end{equation}
where $\hat{P}(\ket{x})=\ket{x}\bra{x}$ and the phase $\theta^a_k$ is random and kept secret. We can imagine that Alice introduces a local ancillary single-mode system $A'_k$ to perform the above procedure equivalently. Alice first chooses a random phase $\theta^a_k \in [0,2\pi)$ and prepares an extended state
\begin{equation}
    \label{eqc}
    \ket{\varphi}_{A'_kA_k} = \frac{1}{\sqrt{2}} \big(\ket{0} \ket{0} + \ket{1} \ket{e^{i\theta^a_k}\sqrt{\mu}} \big)_{A'_kA_k}.
\end{equation}
Then Alice measures the ancillary system $A'_k$ in the basis $\{\ket{s}\}_{s\in\mathbb{Z}_2}$, assigns the local bit as $a_k=0$ or $1$ when the result state is $\ket{0}$ or $\ket{1}$, and sends the system $A_k$ to Charlie. Similarly, Bob could prepare the extended states $\ket{\varphi}_{B'_kB_k}$ by introducing the ancillary system $B'_k$, and perform the same as Alice. 

\begin{figure*}[t]
    \centering
    \includegraphics[width=0.6\textwidth]{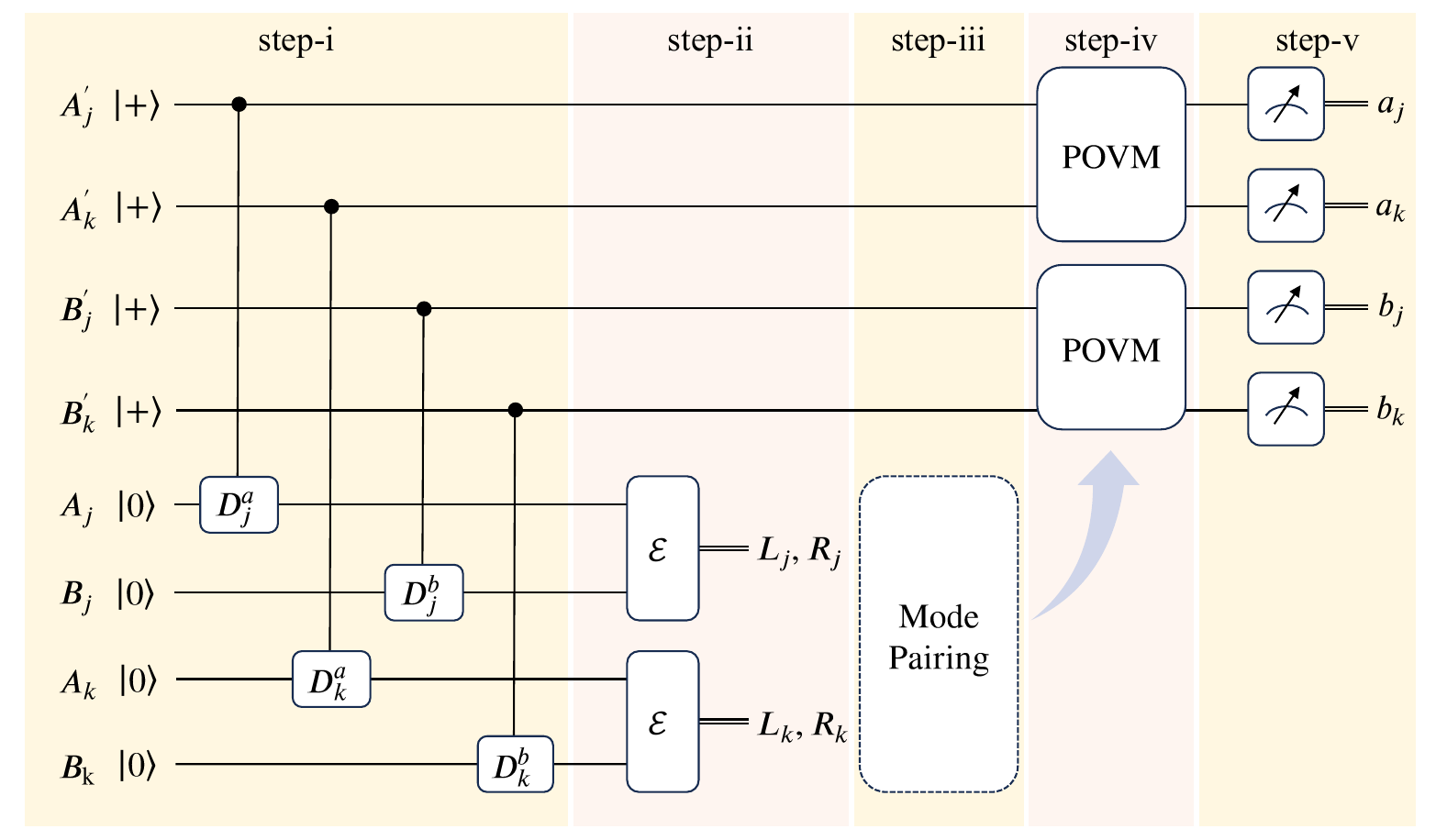}
    \caption{Diagram of the entanglement scheme for MP-QKD. Here we take the pairs in rounds $j$ and $k$ for example, and other rounds are omitted. In step-i, Alice prepares the states $\ket{+}_{A'_k}\ket{0}_{A_k}$ and evolves the states with controlled gate $D_k^a = D(e^{i\theta_k^a} \sqrt{\mu})$ in the $k$-th round, where $D(\gamma) = \exp(\gamma d^\dag - \gamma^* d)$, $D(\gamma)\ket{0} = \ket{\gamma}$, $\gamma\in\mathbb{C}$, $d^\dag$ and $d$ are the creation and annihilation operators. The result state after step-i is $\ket{\varphi}_{A'_kA_k}$ defined in Eq. (\ref{eqc}). Bob prepares the state in the same way independently. Charlie performs the measurement on Alice and Bob's systems $A_kB_k$ in step-ii and announces the results $L_k$ and $R_k$. Then Alice and Bob pair the rounds in step-iii according to the measurement results. They perform the POVM on the paired ancillary systems to establish the entanglement in step-iv, and measure the ancillary systems in the basis $\{\ket{st}\}_{s,t\in\mathbb{Z}_2}$ to obtain the random bits in step-v. The postprocessing process in step-vi and vii are omitted.}
    \label{fig1}
\end{figure*}

Note that the state in Eq. (\ref{eqc}) is assumed as a maximally entangled state in the ancillary and actual systems for every round. But the states in different rounds are independent. We start with these entangled states and consider that Alice and Bob's measurements are performed after Charlie's announcement. A key point is that Alice and Bob could perform based on Charlie's announced detection results. This enables the free pairing in the entanglement scheme. The free-pairing entanglement scheme for MP-QKD is presented in Box 3 and illustrated in Fig. \ref{fig1}. We will analyze the entanglement scheme in detail and prove the security in Sec. \ref{proofofes}.

\begin{tcolorbox}[title = {Box 3: Entanglement Scheme for MP-QKD}]

    (i) State preparation. In the $k$-th round, Alice chooses a random phase $\theta^a_k \in [0,2\pi)$ and prepares $\ket{\varphi}_{A'_kA_k}$. Similarly, Bob chooses $\theta^b_k \in [0,2\pi)$ and prepares $\ket{\varphi}_{B'_kB_k}$.

    (ii) Measurement. Alice and Bob send the systems $A_k$ and $B_k$ to Charlie, and Charlie performs the same as step-2 in Box 1.

    (iii) Mode pairing. Same as step-3 in Box 1.

    (iv) Basis sifting. Alice (Bob) conducts a POVM on the composite systems $A'_jA'_k$ ($B'_jB'_k$) with three operators,
    \begin{equation}
        \begin{aligned}
            \label{eq6}
            M_{0} &= \ket{00}\bra{00},\\
            M_{1} &= \ket{01}\bra{01} + \ket{10}\bra{10},\\
            M_{2} &=\ket{11}\bra{11}.
        \end{aligned}
    \end{equation}
    They independently assign it as '0', $Z$ or $X$ when the measurement result is $0$, $1$ or $2$, respectively. Others are the same as step-4 in Box 1.

    (v) Key mapping. Alice and Bob announces the phase difference $\delta^{a}_{jk}$ and $\delta^{b}_{jk}$. When both are $Z$ basis, Bob evolves his ancillary systems $B'_jB'_k$ with operator $U_{\delta^{a}_{jk}-\delta^{b}_{jk}}$, where $U_{\delta}$ is a phase gate and is defined as
    \begin{equation}
        U_{\delta} = \hat{P}(\ket{01}) + e^{i\delta} \hat{P}(\ket{10}).
    \end{equation}
    The phase gate is used to align the phase of Bob's phase difference to ensure its consistency with Alice's. Then Alice (Bob) measures the composite systems $A'_jA'_k$ ($B'_jB'_k$) in the basis $\{\ket{st}\}_{s,t\in\mathbb{Z}_2}$ and obtains the result $a_j,a_k$ ($b_j,b_k$) in $Z$ basis. They set their raw key bits the same as step-5 in Box 1.

    (vi)-(vii) Same as step-6 to 7 in Box 1.

\end{tcolorbox}

We observe that the POVM in step-iv plays a crucial role in establishing the entanglement between the composite systems $A'_jA'_k$ ($B'_jB'_k$) as discussed below.

\subsection{Security Proof of Entanglement Scheme for MP-QKD}
\label{proofofes}

We analyze the security of the entanglement scheme for MP-QKD in the following. In each round, the composite prepared state in step-i is $\ket{\varphi}_{A'_kA_k} \ket{\varphi}_{B'_kB_k}$ with random phase $\theta_k^a$ and $\theta_k^b$. Then Charlie will perform the measurement and announce the results $L_k, R_k \in\mathbb{Z}_2$ for every round in step-ii. In step-iii, the rounds $j$ and $k$ are paired according to Charlie's measurement results. After the POVM in step-vi, the composite state in rounds $j$ and $k$ will collapse into different states corresponding to Eve's measurement results $\chi \triangleq (L_j,R_j,L_k,R_k) \in\mathbb{Z}^4_2$ and Alice and Bob's POVM results $m_a,m_b\in\mathbb{Z}_3$ as
\begin{equation}
    \label{eqi}
    \begin{aligned}
        \rho'_{jk} = \sum_{\chi\in\mathbb{Z}^4_2;m_a,m_b\in\mathbb{Z}_3} \hat{P} \Big(&E_{\chi} \ket{\varphi_{m_a,\theta_j^a,\theta_k^a}}_{A'A_{jk}} \otimes \ket{\varphi_{m_b,\theta_j^b,\theta_k^b}}_{B'B_{jk}}\Big),
    \end{aligned}
\end{equation}
where $E_{\chi}$ is the operator on systems $A_{jk}B_{jk}$ corresponding $\chi\in\mathbb{Z}^4_2$. The subscripts $A'_{jk}A_{jk}$, $A_{j}A_{k}$ and $A'_{j}A'_{k}$ for Alice are abbreviated as $A'A_{jk}$, $A_{jk}$ and $A'_{jk}$, and it is the same for Bob's systems. Here, the (unnormalized) state $\ket{\varphi_{m,\theta_1,\theta_2}}$ in Eq. (\ref{eqi}) is defined as
\begin{equation}
    \label{eq1}
    \begin{aligned}
        \ket{\varphi_{0,\theta_1,\theta_2}} &= \frac{1}{2} \ket{00} \ket{00},\\
        \ket{\varphi_{1,\theta_1,\theta_2}} &= \frac{1}{2} \sum_{t\in\mathbb{Z}_2} \ket{t} \ket{\bar{t}} \ket{e^{i\theta_1}\sqrt{t\mu}} \ket{e^{i\theta_2}\sqrt{\bar{t}\mu}},\\
        \ket{\varphi_{2,\theta_1,\theta_2}} &= \frac{1}{2} \ket{11} \ket{e^{i\theta_1}\sqrt{\mu}} \ket{e^{i\theta_2}\sqrt{\mu}},
    \end{aligned}
\end{equation}
where the state $\ket{e^{i\theta}\sqrt{t\mu}}$ denotes the vacuum state $\ket{0}$ when $t=0$. The normalized forms of these states in Eq. (\ref{eq1}) can be shown as
\begin{equation}
    \begin{aligned}
        \label{eq5}
        \ket{\varphi_{0,\theta_1,\theta_2}^\prime} &= 2 \ket{\varphi_{0,\theta_1,\theta_2}}, \\
        \ket{\varphi_{1,\theta_1,\theta_2}^\prime} &= \sqrt{2} \ket{\varphi_{1,\theta_1,\theta_2}}, \\
        \ket{\varphi_{2,\theta_1,\theta_2}^\prime} &= 2 \ket{\varphi_{2,\theta_1,\theta_2}}.
    \end{aligned}
\end{equation}
In this way, the relevance is established between systems $A_{jk}$ ($B_{jk}$) and $A'_{jk}$ ($B'_{jk}$) explained as follows.

In step-v, Alice and Bob will announce the phase difference $\delta^a_{jk}$ and $\delta^b_{jk}$, i.e., $\delta \triangleq \theta_1 - \theta_2$ for Eq. (\ref{eq1}). Though the detailed phases $\theta_j^a$ ($\theta_j^b$) and $\theta_k^a$ ($\theta_k^b$) are kept secret, they can not be seen as two independent and random variables due to the announcement of the phase difference. If we re-express the last two states in Eq. (\ref{eq5}) as $\ket{\varphi_{1,\theta + \delta,\theta}^\prime}$ and $\ket{\varphi_{2,\theta + \delta,\theta}^\prime}$, then $\delta$ is announced and fixed while $\theta$ is random and secret. Hence, the states are mixed corresponding to the variable $\theta$ and can be expressed as below, which is analyzed in App. \ref{sup_det_cal}
\begin{equation}
    \begin{aligned}
        \label{eqk}
        \rho_1 &= \frac{1}{2\pi} \int_{0}^{2\pi} \hat{P} \big(\ket{\varphi_{1,\theta + \delta,\theta}^\prime} \big) d\theta = \sum_{n\in\mathbb{N}} p_{\mu,n} \hat{P} \big( \ket{\varphi_{n,\delta}} \big),\\
        \rho_2 &= \frac{1}{2\pi} \int_{0}^{2\pi} \hat{P} \big(\ket{\varphi_{2,\theta + \delta,\theta}^\prime} \big) d\theta = \sum_{n\in\mathbb{N}} p_{2\mu,n} \hat{P} \big( \ket{\phi_{n,\delta}} \big),\\
    \end{aligned}
\end{equation}
where the probability $p_{\tau,n} = e^{-\tau} \tau^n / n!$ is the Poisson distribution probability, and the states $\ket{\varphi_{n,\delta}}$ and $\ket{\phi_{n,\delta}}$ are defined as
\begin{equation}
    \label{eq2}
    \begin{aligned}
        \ket{\varphi_{n,\delta}} &= \frac{1}{\sqrt{2}} \big( \ket{01} \ket{0n} + e^{in\delta} \ket{10} \ket{n0} \big),\\
        \ket{\phi_{n,\delta}} &= \ket{11} \ket{\omega_{n,\delta}}.
    \end{aligned}
\end{equation}
Here, the two-mode $n$-photon state $\ket{\omega_{n,\delta}}$ is defined as
\begin{equation}
    \label{eq7}
    \ket{\omega_{n,\delta}} = \frac{1}{\sqrt{2^n}} \sum_{r=0}^n \sqrt{C_n^r} e^{ir\delta} \ket{r,n-r}.
\end{equation}
Eq. (\ref{eqk}) means that it is equivalent that Alice (Bob) has prepared the mixtures of $\ket{\varphi_{n,\delta}}$ and $\ket{\phi_{n,\delta}}$ when the POVM results are $1$ and $2$, respectively. Note that the states $\ket{\varphi_{n,\delta}}$ and $\ket{\phi_{n,\delta}}$ are independent of the intensity $\mu$, which is the basis of the decoy-state method. 

To see the establishment of the entanglement clearly, we add the subscripts for the state $\ket{\varphi_{n,\delta}}$ defined in Eq. (\ref{eq2}) as
\begin{equation}
    \begin{aligned}
        \label{eq8}
        \ket{\varphi_{n,\delta}}_{A'A_{jk}} =& \frac{1}{\sqrt{2}} \big(\ket{01}_{A'_jA'_k} \ket{0n}_{A_jA_k} + e^{in\delta} \ket{10}_{A'_jA'_k} \ket{n0}_{A_jA_k} \big).
    \end{aligned}
\end{equation}
We can see that the systems $A'_jA'_k$ and $A_jA_k$ are in a maximally entangled state for every element $\ket{\varphi_{n,\delta}}_{A'A_{jk}}$. This means the entanglement is established between the composite systems $A_{jk}$ and $A'_{jk}$ when the POVM result is 1, which can be used to generate key bits. Besides, the state $\ket{\varphi_{n,\delta}}$ defined in Eq. (\ref{eq2}) is shown as below by adding the subscripts
\begin{equation}
    \label{eq9}
    \ket{\phi_{n,\delta}}_{A'A_{jk}} = \ket{11}_{A'_jA'_k} \ket{\omega_{n,\delta}}_{A_jA_k}.
\end{equation}
The systems $A_j$ and $A_k$ are entangled as shown in Eq. (\ref{eq7}). Especially, the single-photon state
\begin{equation}
    \label{eq10}
    \ket{\omega_{1,\delta}}_{A_jA_k} = \frac{1}{\sqrt{2}} (\ket{01}_{A_jA_k} + e^{i\delta} \ket{10}_{A_jA_k}),
\end{equation}
is a maximally entangled state. This means the entanglement between the systems $A_{j}$ ($B_{j}$) and $A_{k}$ ($B_{k}$) is established when the POVM result is 2, which can be used to estimate the phase error rate.

In step-v, the phase differences for Alice and Bob are $\delta \triangleq \delta_{jk}^a$ and $\delta_{jk}^b$, which may be different. Then Bob evolves his ancillary systems in $Z$ basis with operator $U_{\delta_{jk}^a - \delta_{jk}^b}$ according to the announced phase differences. Hence the form of the single-photon states for Alice and Bob in $Z$ basis are both $\ket{\varphi_{1,\delta}}$. By applying the tagging method \cite{gottesman2004RN266}, the secret key rate can be obtained as expressed in Eq. (\ref{rate}). The key is to estimate the fraction of effective single-photon pairs $q_{11}^z$ among $Z$ basis and its phase error rate $e_{11}^x$. The fraction $q_{11}^z$ can be estimated with the decoy-state method \cite{zeng2022RN816}. In the following, we define and analyze how to estimate the phase error rate of the effective single-photon pairs, which characterizes the information leakage.

The single-photon entanglement state $\ket{\varphi_{11,\delta}}$ can be reformulated as
\begin{equation}
    \label{eq3}
    \ket{\varphi_{1,\delta}} = \frac{1}{\sqrt{2}}\big(\ket{\omega_{1,0}}\ket{\omega_{1,\delta}} + \ket{\omega_{1,\pi}}\ket{\omega_{1,\delta+\pi}}\big).
\end{equation}
Consider that Alice and Bob have virtually measured the ancillary systems of the single-photon state $\ket{\varphi_{1,\delta}}$ in $X_0 = \{\ket{\omega_{1,0}},\ket{\omega_{1,\pi}}\}$ basis but not $\{\ket{st}\}_{s,t\in\mathbb{Z}_2}$ in step-v. We note that $\{\ket{01},\ket{10}\}$ and $X_0$ are mutually unbiased bases, and the error rate in $X_0$ basis can be defined as the phase error rate to characterize the information leakage. Due to the entanglement property, the emitted states are $\ket{\omega_{1,\delta}}$ and $\ket{\omega_{1,\delta+\pi}}$ when the local states are $\ket{\omega_{1,0}}$ and $\ket{\omega_{1,\pi}}$, respectively. Note that the pairs imply that $L_j \oplus R_j = L_k \oplus R_k = 1$. Consider all the pairs when Alice and Bob emit the states $\ket{\omega_{1,\delta}}$ or $\ket{\omega_{1,\delta+\pi}}$, the phase error rate $e_{11}^x(\delta)$ is defined as the fraction of two cases: $L_j \neq L_k$ while Alice and Bob emit the same states; $L_j = L_k$ while Alice and Bob emit different states. The phase difference $\delta$ is random and the overall phase error rate can be obtained using the convexity of the binary entropy function as
\begin{equation}
    e_{11}^x = \frac{1}{2\pi} \int_{0}^{2\pi} e_{11}^x(\delta) d\theta
\end{equation}
Note that the single-photon states emitted in $X$ basis is just the state $\ket{\omega_{1,\delta}}$ as shown in Eq. (\ref{eq2}). In this way, we can estimate the phase error rate when Alice and Bob emit the single-photon state $\ket{\omega_{1,\delta_{jk}^a}} \ket{\omega_{1,\delta_{jk}^b}}$ by sifting the pairs satisfying $\delta_{jk}^a = \delta_{jk}^b \mod \pi$ and performing the decoy-state method \cite{zeng2022RN816}.

\subsection{Reduction to Prepare-and-Measure Scheme for MP-QKD}

We prove the security of MP-QKD by demonstrating the equivalence of the prepare-and-measure scheme and the free-pairing entanglement scheme for MP-QKD. The key is to prove the commutativity of Alice and Bob's operations and Charlie's measurement. Below, we present the reduction of the entanglement scheme to the prepare-and-measure scheme progressively through three intermediate entanglement schemes in Boxes 4, 5, and 6. 

In the entanglement scheme for MP-QKD, Alice first prepares the entangled state in Eq. (\ref{eqc}). We note that this entanglement is only assumed for every single round $k$. But the quantum states in different rounds are uncorrelated and assumed to be independent and identically distributed. Then Alice tries to establish entanglement between two paired rounds to extract the key bits. This is achieved by introducing the POVM defined in step (iv) and the announcement of phase differences in step (v). By performing POVM, the states in paired rounds will collapse to the state defined in Eq. (\ref{eqi}) with Alice and Bob's states defined in Eq. (\ref{eq1}). Then it shows that the states are equivalent to the states $\rho_1$ and $\rho_2$ defined in Eq. (\ref{eqk}). As analyzed in Eqs. (\ref{eq8}), (\ref{eq9}), and (\ref{eq10}), the entanglement is established when performing the POVM and obtaining the measurement results 1 and 2.

However, the POVM must be removed because there is no such operation in the prepare-and-measure scheme. The common method to prove equivalence is demonstrating that POVM can be advanced in the state preparation step. However, the paired location $j$ and $k$ in the entanglement scheme for MP-QKD are determined by the measurement results announced by Charlie in step-ii. This means Alice and Bob could not perform the POVM in step-i ahead because they are unable to predict the measurement results. Besides, the evolution and measurements in step-v should also be removed. We provide a detailed analysis of how this can be accomplished in the following.

In the entanglement scheme for MP-QKD, Alice and Bob perform the POVM in step-iv, modulate the phase and perform measurement in the basis $\{\ket{st}\}_{s,t\in\mathbb{Z}_2}$ on the ancillary systems in step-v. The phase evolution with operator $U_{\delta^{a}_{jk}-\delta^{b}_{jk}}$ has no physical observability on the last measurement in the basis $\{\ket{st}\}_{s,t\in\mathbb{Z}_2}$, which can be eliminated directly. Note that the phase evolution is performed only in $Z$ basis, hence the elimination will not affect the estimation of the phase error rate with $X$ basis. Since these operations are done continuously after Charlie's announcement, there is no difference if we combine them into a single operation as long as we can infer the intermediate results (i.e., the POVM results). The combined operation is just the measurement in step (v). When measuring the ancillary systems in two steps, the measurement result state is $\ket{00}$, $\ket{11}$, and $\ket{01}$ or $\ket{10}$ in the basis $\{\ket{st}\}_{s,t\in\mathbb{Z}_2}$ only when the POVM result is 0, 2, and 1, respectively. Therefore, we can infer the POVM results from the results measured in the basis $\{\ket{st}\}_{s,t\in\mathbb{Z}_2}$. Therefore, the entanglement scheme for MP-QKD in Sec. \ref{entanglement} is equivalent to the following  \textit{entanglement scheme for MP-QKD II} in Box 4 by eliminating the POVM and phase evolution and advancing the measurement in the basis $\{\ket{st}\}_{s,t\in\mathbb{Z}_2}$ from step-v to step-iv:

\begin{tcolorbox}[title = {Box 4: Entanglement Scheme for MP-QKD II}]
 
    (i')-(iii') Same as step-i to iii in Box 3.

    (iv') Basis sifting. Aice (Bob) measures the composite systems $A'_jA'_k$ ($B'_jB'_k$) in the basis $\{\ket{st}\}_{s,t\in\mathbb{Z}_2}$ and obtains the results $a_j$, $a_k$ ($b_j$, $b_k$). Alice assigns it as '0', $Z$ or $X$ when the results $a_j=a_k=0$, $a_j+a_k = 1$ or $a_j=a_k=1$. And Bob assigns the basis in the same way. Others are the same as step-4 in Box 1.

    (v') Key mapping. They set their key bits the same as step-5 in Box 1.
    
    (vi')-(vii') Same as step-6 to 7 in Box 3.

\end{tcolorbox}

In the step-iv', the measurement on the systems $A'_{jk}$ ($B'_{jk}$) in the basis $\{\ket{st}\}_{s,t\in\mathbb{Z}_2}$ is just the measurement on the systems $A'_j$ and $A'_k$ ($B'_j$ and $B'_k$) in the basis $\{\ket{s}\}_{s\in\mathbb{Z}_2}$, separately. Therefore, \textit{entanglement scheme for MP-QKD II} is equivalent to the following \textit{entanglement scheme for MP-QKD III} in Box 5:

\begin{tcolorbox}[title = {Box 5: Entanglement Scheme for MP-QKD III}]

    (i'')-(iii'') Same as step-i to iii in Box 3.

    (iv'') Basis sifting. Aice (Bob) measures the composite systems $A'_j$ and $A'_k$ ($B'_j$ and $B'_k$) in the basis $\{\ket{s}\}_{s\in\mathbb{Z}_2}$ separately and obtains the results $a_j$, $a_k$ ($b_j$, $b_k$). Others are the same as step-iv' in Box 4.

    (v'')-(vii'') Same as step-5 to 7 in Box 3.

\end{tcolorbox}

The measurement on every ancillary systems $A'_k$ and $B'_k$ in the basis $\{\ket{s}\}_{s\in\mathbb{Z}_2}$ and Charlie's measurement on the systems $A_k$ and $B_k$ are acted on different systems and independent of each other. This means Alice and Bob's measurements and Charlie's measurement are commute. Thus, Alice could prepares the states $\ket{\varphi}_{A'_kA_k}$ in Eq. (\ref{eqc}), measures the system $A'_k$ in the basis $\{\ket{s}\}_{s\in\mathbb{Z}_2}$ and then sends the system $A_k$ to Charlie at once in the state-preparation step. Hence, the \textit{entanglement scheme for MP-QKD III} is equivalent to the following \textit{entanglement scheme for MP-QKD IV} in Box 6:

\begin{tcolorbox}[title = {Box 6: Entanglement Scheme for MP-QKD IV}]

    (1') State preparation. Same as step-i in Box 3. Then Alice (Bob) measures the system $A'_k$ ($B'_k$) in the basis $\{\ket{s}\}_{s\in \mathbb{Z}_2}$ and obtains the result $a_k$ ($b_k$).

    (2')-(7') Same as step-2 to 7 in Box 1.

\end{tcolorbox}

When measuring the ancillary system $A'_k$ ($B'_k$), the quantum states in the actual system $A_k$ ($B_k$) are collapsed accordingly due to the entanglement. It is equivalent to preparing the collapsed quantum states in the actual systems directly. The preparation of the systems $A'_k$ and $B'_k$ can be removed. The MP-QKD protocol is equivalent to the \textit{entanglement scheme for MP-QKD IV} in Box 6. In this way, we complete the reduction of the entanglement scheme in Box 3 to the prepare-and-measure scheme in Box 1, thereby proving the security of the latter.

\section{Analysis of Pairing Strategy}
\label{analysis}

In this section, we try to optimize the pairing strategy based on the entanglement model. In Box 2, the pairing strategy is based only on the announced measurement results. When the decoy-state method is employed, there will be some mismatched pairs useless for coding. For example, the vacuum + weak decoy-state method utilizes three intensities $\{\mu,\nu,0\}$ ($0 < \nu < \mu < 1$) as presented in App. \ref{decoy_state_protocol}. First, there are mismatched rounds where Alice and Bob's intensities are $(\mu,\nu)$ or $(\nu,\mu)$, respectively. Second, the mismatched pairs will be formed even if the rounds are matched. Only the paris when Alice and Bob's intensities are $(0,\mu)$ or $(\mu,0)$ (i.e., $Z$ basis) can be used to generate raw key bits. Below we analyze how to reduce the mismatched pairs to improve the pairing efficiency $r_z$.

The improved pairing strategy for decoy-state MP-QKD is present in Box 7. Here, we define the label $\mathcal{L}_i^a$ ($\mathcal{L}_i^b$) of every round for Alice (Bob), which is $0$ if the intensity is $\mu$ or $0$, or $1$ if the intensity is $\nu$. Based on the labels $\mathcal{L}_i^a,\mathcal{L}_i^b$ and Charlie's announced results $C_i = L_i \oplus R_i$ in every round, they perform the pairing strategy. The pairing strategy can be divided into two stages. In the first stage, Alice and Bob sift the rounds when their labels are the same ($\mathcal{L}_i^a$ = $\mathcal{L}_i^b$) and the detection is effective ($C_i=1$). This filters out the mismatched rounds. The sifted rounds are independent since this is performed independently for every round. In the second stage, they perform the pairing. If they pair adjacent sifted rounds in the maximal pairing interval, there will be some mismatched pairs where $\mathcal{L}_i^a \neq \mathcal{L}_i^a$ which is useless for raw key bits. In fact, they could avoid the mismatched pairs as the labels are announced. If the adjacent sifted pairs are mismatched, they discard the front round, regard the rear round as a new front round and try to pair it with another rear round. In this way, the mismatched pairs are avoided and the pairing efficiency $r_z$ can be improved. The general idea that the pairing strategy is secure is that the entanglement can still be distilled in pairs with the announcement of labels and abandonment of mismatched rounds. Detailed proof is given in App. \ref{decoy_state_protocol}.

\begin{tcolorbox}[title = {Box 7: Improved Pairing Strategy}]

    \textbf{Input}: Alice and Bob's labels $\mathcal{L}_i^a,\mathcal{L}_i^b$, and Charlie's announced detection results $C_i = L_i \oplus R_i$ for $i=1$ to $N$; maximal pairing interval $l$.

    \textbf{Output}: $K$ pairs, $(F_k,R_k)$ for the $k$-th pair for $k=1$ to $K$.

    \textbf{Initialization}: $k=1$, $f=0$, $d=0$; $D_i = -1$ for $i=1$ to $N$.

    \textbf{for} $i\in[N]$ \textbf{do}

    \quad \textbf{if} $(\mathcal{L}_i^a,\mathcal{L}_i^b) = (0,0)$ \textbf{then}

    \quad\quad $D_i = 0$.

    \quad \textbf{else if} $(\mathcal{L}_i^a,\mathcal{L}_i^b) = (1,1)$ \textbf{then}

    \quad\quad $D_i = 1$.

    \quad \textbf{end if}

    \textbf{end for}

    \textbf{for} $i\in[N]$ \textbf{do}

    \quad \textbf{if} $C_i=1$ and $D_i \neq -1$ \textbf{then}

    \quad\quad \textbf{if} $f = 0$ \textbf{then}

    \quad\quad\quad $F_k\leftarrow i$; $f\leftarrow 1$; $d\leftarrow D_i$.

    \quad\quad \textbf{else then}

    \quad\quad\quad \textbf{if} $i-F_k\leq l$ and $D_i = d$ \textbf{then}

    \quad\quad\quad\quad $R_k\leftarrow i$; $k\leftarrow k+1$; $f\leftarrow 0$.

    \quad\quad\quad \textbf{else} \textbf{then}

    \quad\quad\quad\quad $F_k\leftarrow i$; $d\leftarrow D_i$.

    \quad\quad\quad \textbf{end if}

    \quad\quad \textbf{end if}

    \quad \textbf{end if}

    \textbf{end for}

\end{tcolorbox}

\begin{figure}[t]
    \centering
    \includegraphics[width=0.47\textwidth]{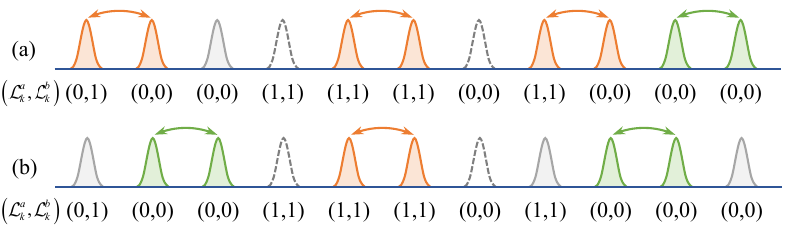}
    \caption{The example of the pairing results with pairing strategies in Boxes 2 and 7. Here we assume the maximal pairing interval $l=2$ for simplicity. The solid and dashed pulses correspond to the effective rounds when $L_k \oplus R_k = 1$ and other ineffective rounds. The gray solid pulses denote the unpaired rounds, and the green and red solid pulses denote the paired rounds. Note only those green pairs can be used to generate raw key bits. (a): The pairing results with the pairing strategy in Box 2, which pairs adjacent effective rounds directly. (b): The pairing results with the pairing strategy in Box 7. In the first stage, those rounds where $\mathcal{L}_k^a \neq \mathcal{L}_k^b$ are filtered out (e.g., the first round). In the second stage, the sifted effective pairs are paired when $\mathcal{L}_j^a = \mathcal{L}_k^b$. Hence the eighth round is unpaired as it does not match the ninth. The number of all pairs in (b) is less than that in (a) but the pairs for raw key bits are more in (b).}
    \label{fig_pairing}
\end{figure}

In Fig. \ref{fig_pairing}, we show a simple example to compare the two pairing strategies in Boxes 2 and 7. We analyze and compare the pairing efficiency $r_z$ of these two strategies, which determines the secret key rate in Eq. (\ref{rate}). Define $q$ as the probability of an effective round (i.e., $C_i = 1$). Let $q_0$ and $q_1$ denote the probabilities of an effective round and $\mathcal{L}_i^a = \mathcal{L}_i^b =$ 0 and 1, respectively. Since there will be some mismatched rounds where $\mathcal{L}_i^a \neq \mathcal{L}_i^b$, the probabilities satisfy $q \geq q' \triangleq q_0 + q_1$. These probabilities can be expressed as
\begin{equation}
    \begin{aligned}
        q &= \sum_{\tau_a,\tau_b \in \{\mu,\nu,0\}} p_{\tau_a} p_{\tau_b} q(\tau_a,\tau_b),\\
        q_0 &= \sum_{\tau_a,\tau_b \in \{\mu,0\}} p_{\tau_a} p_{\tau_b} q(\tau_a,\tau_b),\\
        q_1 &= p_{\nu}^2 q(\nu,\nu),
    \end{aligned}
\end{equation}
where $q(\tau_a,\tau_b)$ is the probability of an effective round when Alice and Bob's intensities are $\tau_a$ and $\tau_b$, respectively.

In Box 2, the pairing is performed based only on the announced measurement results within the maximal interval $l$ and the pairing efficiency of all pairs $r$ can be calculated as \cite{zeng2022RN816}
\begin{equation}
    r = \bigg[ \frac{1}{q} + \frac{1}{q} \frac{1}{1 - (1-q)^l} \bigg]^{-1}.
\end{equation}
And the proportion of $Z$ basis among all pairs is
\begin{equation}
    r_{(z)} = \frac{2}{q^2} p_0^2 p_\mu^2 q(0,\mu) q(\mu,0).
\end{equation}
Hence the pairing efficiency of $Z$ basis is $r_z = rr_{(z)}$ in Box 2.

In Box 7, Alice and Bob will first filter out rounds with $\mathcal{L}_i^a \neq \mathcal{L}_i^b$ and use only the sifted effective rounds for pairing. As analyzed in App. \ref{decoy_state_protocol}, the pairing efficiency of all pairs is
\begin{equation}
    \label{eq4}
    r' = \bigg[ \frac{1}{q'} + \frac{q'}{q_0^2 + q_1^2} \frac{1}{1 - (1 - q')^l} \bigg]^{-1}.
\end{equation}
And the proportion of $Z$ basis among all pairs is
\begin{equation}
    r'_{(z)} = \frac{2}{q_0^2 + q_1^2} p_0^2 p_\mu^2 q(0,\mu) q(\mu,0).
\end{equation}
We obtain the pairing efficiency of $Z$ basis as $r'_z = r'r'_{(z)}$ in Box 7.

We compare the two pairing strategies with simulation. As the $Z$ basis is formed with adjacent effective rounds, the difference between the two pairing strategies is just the efficiency and other parameters are not affected. Hence the ratio of the secret key rate is equal to the ratio of $r_z$ and $r'_z$. To simulate the ratio, we give the simulation method of the probability $q(\tau_a,\tau_b)$ as
\begin{equation}
    \begin{aligned}
        q(\tau_a,\tau_b) = \frac{1}{2\pi} \int_{0}^{2\pi} & \Big\{ e^{-\tau_L(\tau_a,\tau_b,\theta)} (1 - p_d) \times [1 - e^{-\tau_R(\tau_a,\tau_b,\theta)} (1 - p_d)] \\
        + & [1 - e^{-\tau_L(\tau_a,\tau_b,\theta)} (1 - p_d)] \times  e^{-\tau_R(\tau_a,\tau_b,\theta)} (1 - p_d) \Big\} d\theta,
    \end{aligned}
\end{equation}
where
\begin{equation}
    \begin{aligned}
        \tau_L(\tau_a,\tau_b,\theta) &= \frac{\eta}{2}(\tau_a\eta_a + \tau_b\eta_b) + \eta\sqrt{\tau_a\eta_a\tau_b\eta_b} \cos\theta,\\
        \tau_R(\tau_a,\tau_b,\theta) &= \frac{\eta}{2}(\tau_a\eta_a + \tau_b\eta_b) - \eta\sqrt{\tau_a\eta_a\tau_b\eta_b} \cos\theta.\\
    \end{aligned}
\end{equation}
Here, $\eta$ and $p_d$ are the detection efficiency and dark count probability of the single-photon detectors (SPDs), $\eta_a$ and $\eta_b$ are the transmission efficiency from Alice and Bob to Charlie.

\begin{figure}[t]
    \centering
    \includegraphics[width=0.45\textwidth]{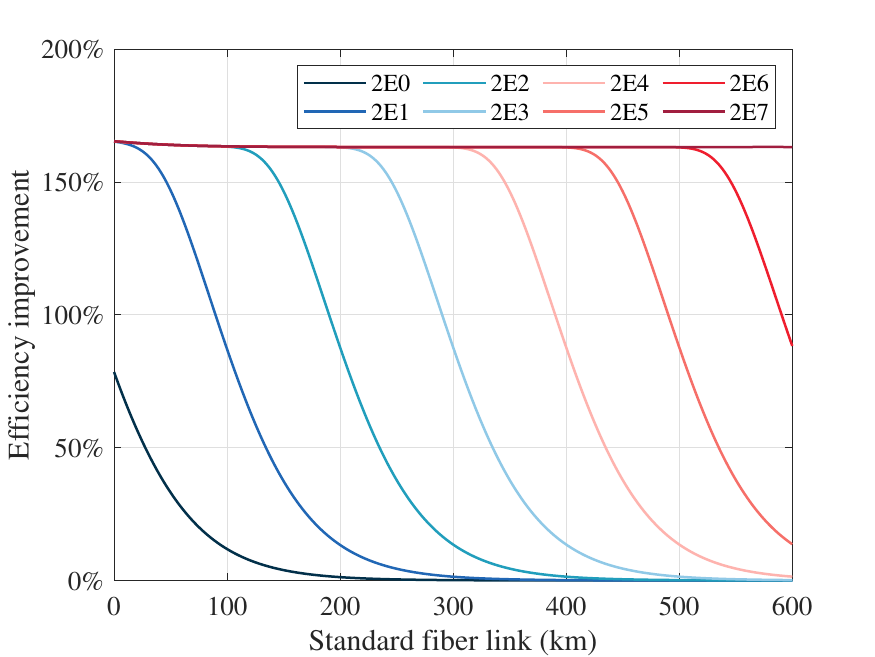}
    \caption{The improvement of the pairing efficiency defined as $(r'_z - r_z) / r_z$ with different pairing interval $l$.}
    \label{fig_ratio}
\end{figure}

We set the parameters of SPDs as $\eta = 80\%$ and $p_d = 1$E$-9$, the intensities $\mu=0.5$ and $\nu=0.1$, and the probabilities $\Pr[\mathcal{L}_k^{a(b)} = 0 (1)] = 0.5$ fairly. We simulate the pairing efficiency $r_z$ and $r'_z$ of the pairing strategies in Boxes 2 and 7, respectively. The percentage increase in efficiency defined as $(r'_z - r_z) / r_z$ under different pairing interval $l$ is shown in Fig. \ref{fig_ratio}. The results show that the improvement is determined by the distance and pairing interval, with significant gains observed at short distances or with larger pairing intervals. The improvement remains steady at first and exhibits a declining trend after a specific distance. Moreover, as the pairing interval increases, the point at which the decline starts becomes farther. For example, the pairing efficiency is improved by greater than 163\% within 400 km given a reasonable pairing interval $l=2$E5 \cite{zhou2023RN978} and then reduces gradually after 400 km. This decline is due to an insufficient number of effective rounds within the pairing interval. In Fig. \ref{fig_ratio}, there are two extreme cases where $l=2$ and 2E7. It means Alice and Bob only pair adjacent rounds when $l=2$, and the result shows that the improvement is 78\% at 0 km and disappears until about 200 km. It is a large pairing interval when $l=$2E7 and the result shows that the improvement is greater than 163\% between 0 and 600 km. Note that the ratio of the pairing efficiency just equal to the ratio of the secret key rate in the asymptotic case. Therefore, the improved pairing strategy in Box 7 significantly enhances the performance of MP-QKD.

\section{Conclusion}
\label{conclusion}

MP-QKD is a promising protocol that simultaneously achieves practicality and efficiency. Its security has been rigorously proven in previous work by examining the consistency of Alice and Bob's states and Charlie's possible measurement between MP-QKD and the fixed-pairing scheme to establish equivalence, as well as verifying the security of the latter. In this work, we present an entanglement model by proposing a free-pairing entanglement scheme that is equivalent to MP-QKD. In the entanglement scheme, we design a novel POVM to establish entanglement in paired rounds. We prove the commutativity of Alice and Bob's operations and Charlie's measurement, and reduce it to the prepare-and-measure scheme (i.e., MP-QKD). With this model, we can explain why the paired rounds can be decoupled and how the entanglement is established based on Charlie's announced results, which are essential for the security of MP-QKD. This entanglement model provides an intuitive and simple theoretical framework for the security and performance analysis of MP-QKD. As an application, we analyze and optimize the pairing strategy, and successfully improve the secret key rate.

\section*{Acknowledgments}
The research was supported by National Key Research and Development Program of China (2020YFA0-309702); National Natural Science Foundation of China (62101597, U2130205); China Postdoctoral Science Foundation (2021M691536); Natural Science Foundation of Henan Province (202300410532; 202300410534); Anhui Initiative in Quantum Information Technologies.

\appendix

\section{Decoy-state MP-QKD}
\label{decoy_state_protocol}

Here we present the vacuum + weak decoy-state MP-QKD as an example and prove its security with the pairing strategy in Box 7 based on the entanglement model.

\begin{tcolorbox}[title = {Box A1: Decoy-State MP-QKD}]

    (a1) State preparation. In $k$-th round, Alice prepares a coherent state $\ket{e^{i\theta_k^a} \sqrt{\tau}}$ with random phase $\theta^a_k \in [0,2\pi)$ and intensity $\tau \in \{\mu,\nu,0\}$ of probability $p_\tau$. She assigns the label as $\mathcal{L}_k^a = 0$ when she chooses intensities $\mu$ or 0, and as $\mathcal{L}_k^a = 1$ when $\nu$. When $\mathcal{L}_k^a = 0$, she sets the local bit as $a_k$ = 0 or 1 if the intensity is 0 or $\mu$, respectively. Bob performs the same procedure independently.

    (a2) Measurement. Same as step-2 in Box 1.

    (a3) Mode pairing. After repeating the above steps $N$ times, Alice and Bob sift the effective rounds with $L_k \oplus R_k=1$ and announce the labels $\mathcal{L}_k^a,\mathcal{L}_k^b$. They perform the pairing according to the pairing strategy in Box 7.

    (a4) Basis sifting. Alice and Bob independently assign the pairs (indexed by $j$ and $k$) as $Z$ basis if the intensities are $(0,\mu)$ or $(\mu,0)$, as $X$ basis if $(\mu,\mu)$ or $(\nu,\nu)$, or as '0' if $(0,0)$. Others are the same as step-4 in Box 1.

    (a5)-(a7) Same as step-5 to 7 in Box 1.

\end{tcolorbox}

To analyze the decoy-state MP-QKD, we give the following entanglement scheme, which is adapted from Box 3 by introducing the decoy states.

\begin{tcolorbox}[title = {Box A2: Entanglement Scheme for Decoy-State MP-QKD}]

    (ai) State preparation. In $k$-th round, Alice chooses a random phase $\theta^a_k \in [0,2\pi)$ and prepares the following state in the composite systems $A'_kA_k$
    \begin{equation}
        \begin{aligned}
            \ket{\varphi'}_{A'_kA_k} =& \big(\sqrt{p_0} \ket{0}\ket{0} + \sqrt{p_\mu} \ket{1}\ket{e^{i\theta_k^a} \sqrt{\mu}} + \sqrt{p_\nu} \ket{2}\ket{e^{i\theta_k^a} \sqrt{\nu}} \big)_{A'_kA_k}.
        \end{aligned}
    \end{equation}
    Then she measures the auxiliary system $A'_k$ with measurement operators
    \begin{equation}
        \begin{aligned}
            \label{eqs4}
            M'_0 &= \ket{0}\bra{0} + \ket{1}\bra{1},\\
            M'_1 &= \ket{2}\bra{2},
        \end{aligned}
    \end{equation}
    and assigns the label $\mathcal{L}_k^a$ as the measurement result 0 or 1. Bob performs the same procedure independently.

    (aii)-(aiii) Same as step-a2 to a3 in Box 1.

    (aiv) Basis sifting. When the labels of the pair $(\mathcal{L}_j^a,\mathcal{L}_k^a) = (0,0)$, Alice performs the POVM defined in step-iv in Box 3 and assigns it as '0', $Z$ or $X$ when the measurement result is 0, 1, or 2. When the labels of the pair $(\mathcal{L}_j^a,\mathcal{L}_k^a) = (1,1)$, Alice assign it as $X$. Bob performs the same procedure, independently. Others are the same as step-4 in Box 1.

    (av)-(avii) Same as step-v to vii in Box 3.

\end{tcolorbox}

In step-ai, Alice performs the measurement and will obtain the label $\mathcal{L}_k^a$ and the following states
\begin{equation}
    \begin{aligned}
        \label{eqs1}
        \ket{\varphi_0}_{A'_kA_k} &= \frac{\sqrt{p_0} \ket{0}\ket{0} + \sqrt{p_\mu} \ket{1}\ket{e^{i\theta_k^a} \sqrt{\mu}}}{\sqrt{p_0 + p_\mu}},\\
        \ket{\varphi_1}_{A'_kA_k} &= \ket{2}\ket{e^{i\theta_k^a} \sqrt{\nu}},
    \end{aligned}
\end{equation}
with probabilities $p_0 + p_\mu$ and $p_\nu$, respectively. The label $\mathcal{L}_k^a$ takes values 0 and 1 just indicate that the states are $\ket{\varphi_0}_{A'_kA_k}$ and $\ket{\varphi_1}_{A'_kA_k}$, respectively. Hence, the announcement of label $\mathcal{L}_k^a$ in step-aiii will not affect their states $\ket{\varphi_0}$ and $\ket{\varphi_1}$ in Eq. \ref{eqs1}. Bob will obtain his results similarly. Then after Charlie's announcement in step-aii, they announce the labels $\mathcal{L}_k^a$ and $\mathcal{L}_k^b$ for those effective rounds in step-aiii. The state $\ket{\varphi_0}$ is analogic to the state $\ket{\varphi}$ defined in Eq. (\ref{eqc}), which can be used to generate key bits. After the POVM on state $\ket{\varphi_0}$ in step-aiv, the composite state in rounds $j$ and $k$ will collapse to the following states, which is analogy to the state $\rho'_{jk}$ defined in Eq. (\ref{eqi})
\begin{equation}
    \begin{aligned}
        \label{eqs3}
        \tilde{\rho}'_{jk} = \sum_{\chi\in\mathbb{Z}^4_2;m_a,m_b\in\mathbb{Z}_3} \hat{P} \Big(&E_{\chi} \ket{\varphi'_{m_a,\theta_j^a,\theta_k^a}}_{A'A_{jk}} \otimes \ket{\varphi'_{m_b,\theta_j^b,\theta_k^b}}_{B'B_{jk}}\Big),
    \end{aligned}
\end{equation}
where the (unnormalized) states $\ket{\varphi'_{m,\theta_1,\theta_2}}$ are

\begin{equation}
    \begin{aligned}
        \label{eqs2}
        \ket{\varphi'_{0,\theta_1,\theta_2}} &= \frac{2p_0}{p_0 + p_\mu} \ket{\varphi_{0,\theta_1,\theta_2}},\\
        \ket{\varphi'_{1,\theta_1,\theta_2}} &= \frac{2\sqrt{p_0p_\mu}}{p_0 + p_\mu} \ket{\varphi_{1,\theta_1,\theta_2}},\\
        \ket{\varphi'_{2,\theta_1,\theta_2}} &= \frac{2p_\mu}{p_0 + p_\mu} \ket{\varphi_{2,\theta_1,\theta_2}}
    \end{aligned}
\end{equation}
The normalized states of Eq. (\ref{eqs2}) are just the normalized states of Eq. (\ref{eq1}). Hence, the subsequent analysis for entanglement distillation in $Z$ basis is the same as that in Sec. \ref{proofofes}. In $X$ basis, the state $\ket{22} \ket{e^{i\theta_1}\sqrt{\nu}} \ket{e^{i\theta_2}\sqrt{\nu}}$ is equivalent to $\sum_{n\in\mathbb{N}} p_{2\nu,n} \hat{P} (\ket{22} \ket{\omega_{n,\delta}})$ with announced $\delta = \theta_1 - \theta_2$ as analyzed in Eq. (\ref{eqk}), where the single-photon state $\ket{\omega_{1,\delta}}$ can be used to estimate the phase error rate.

From the above analysis, we show that the announcement of the labels $\mathcal{L}_i^a$ and $\mathcal{L}_i^b$ will not affect the security. We analyze how to use this announcement to perform pairing, which is shown in Box 7. We define $\mathsf{S}_i = (\mathcal{L}_i^a,\mathcal{L}_i^b,\theta_i^a,\theta_i^b)$ as the local settings of Alice and Bob in $i$-th round, which determine the prepared states in step-ai as analyzed in Eq. (\ref{eqs1}). The settings are determined independently for every round, which can be expressed as
\begin{equation}
    \label{eqa11}
    \Pr(\mathsf{S}_i \mathsf{S}_j) = \Pr(\mathsf{S}_i) \Pr(\mathsf{S}_j), i \neq j.
\end{equation}
First, Alice and Bob could filter out the rounds when Alice and Bob's labels are different. The remained rounds are still independent since this filtering is performed independently for every round. Suppose the $t$-th round is filtered out (denoted as $\mathsf{S}'_t$), which just mean $\mathsf{S}_t$ takes the special case $\mathcal{L}_t^a \neq \mathcal{L}_t^b$. Note that the dependence between the filtered out rounds and other remaining rounds are still valid, i.e., $\Pr(\mathsf{S}_i \mathsf{S}'_t) = \Pr(\mathsf{S}_i) \Pr(\mathsf{S}'_t)$. Hence the independence of the remained rounds (e.g. $i$ and $j$) can be easily verified as
\begin{equation}
    \Pr(\mathsf{S}_i \mathsf{S}_j | \mathsf{S}'_t) = \Pr(\mathsf{S}_i | \mathsf{S}'_t) \Pr(\mathsf{S}_j | \mathsf{S}'_t).
\end{equation}
Besides, Alice and Bob only consider the rounds when $C_i = 1$. In other words, the ineffective rounds when $C_i = 0$ are filtered out. This filtering is performed regardless of the local settings $\mathsf{S}_k$, which is the same as that in Box 2. Hence the independence of the remained rounds can still be guaranteed. Now if they pair the adjacent remaining rounds in the maximal pairing interval, there still will be some mismatched pairs where the labels of the front and rear rounds are different. As the labels of the rounds are announced, they could avoid the mismatched pairs. If the adjacent remaining pairs are mismatched, they discard the front round, regard the rear round as a new front round and try to pair it with another rear round. We note that when the rounds are discarded due to the mismatch of windows, any additional information is not leaked as the labels are announced in advance. Specifically, the phase is still secret since the phase is not announced for pairing and the rounds are independent as analyzed in Eq. (\ref{eqa11}). Hence Alice and Bob still could obtain the state $\tilde{\rho}'_{jk}$ in Eq. (\ref{eqs3}) to distill entanglement and perform parameter estimation. Besides, there is no correlation between the pairs since a new pairing starts only after the current pairing is completed. To prove this, we suppose the set of rounds $X_k = \{\mathsf{S}_{k_0},\mathsf{S}_{k_1},...,\mathsf{S}_{k_{k'}}\}$ that are tried for the $k$-th pairs, where $k_0 < k_1 < ... < k_{k'}$. Specifically, the rounds in $X_k$ satisfy: Alice and Bob's labels are the same; the detection is effective. In this way, the $k$-th pairs can be directly determined by the set $X_k$ and can be denoted as $(\mathsf{S}_{k_{k'}},\mathsf{S}_{k_{k'-1}})$. Note that $k_{k'} < (k+1)_0$ for the $k$ and $k+1$-th pairs, hence the independence of the sets $X_k$ and $X_{k+1}$ can be obtained. In this way, the $k$ and $k+1$-th pairs are independent. Similarly, the independence of all pairs can be proven.

At last, we analyze the pairing efficiency in Eq. (\ref{eq4}). The front and rear indexes are denoted as $(F_k,R_k)$ for the $k$-th pair. We assume the starting index $S_k$ as the first sifted effective round that is attempted for the $k$-th pair. Note that the $S_k$-th round may be discarded if the window of the next sifted effective round is different or the interval length is larger than the maximal interval $l$. To analyze the pairing efficiency, we define $G_k = S_{k+1} - S_k$ as the interval length between the $k$-th and $k+1$-th starting rounds. By analyzing the expectation value $\mathbb{E} (G_k)$, we can obtain the pairing efficiency as $r' = 1 / \mathbb{E} (G_k)$. First, the interval distance $G_k$ can be divided as $G_k = H_k + L_k$ and hence $\mathbb{E}(G_k) = \mathbb{E}(H_k) + \mathbb{E}(L_k)$, where $L_k = S_{k+1} - R_k$ and $H_k = R_k - S_k$. The variable $L_k$ follows a geometric distribution $\Pr (L_k = d) = (1-q')^{d-1} q'$ for $d\in\mathbb{N^*}$ and the expectation value is $\mathbb{E}(L_k) = 1/q'$. To calculate the expectation value $\mathbb{E}(H_k)$, we define the conditional expectation value $\mathbb{E}(H_k|d)$ when the interval length of $S_k$ and the following sifted effective round is $d$, which can be calculated as
\begin{equation}
    \begin{aligned}
        &\mathbb{E}(H_k|d)
        = \left\{
        \begin{array}{l}
            q_c d + (1 - q_c) [\mathbb{E}(H_k) + d], d\leq l\\
            \mathbb{E}(H_k) + d, d>l,
        \end{array}
      \right.
    \end{aligned}
\end{equation}
where $q_c = (q_0^2 + q_1^2) / q^{\prime 2}$.The probability that the interval length of $S_k$ and the following sifted effective round is $d$ can be given as $\Pr(d) = (1-q')^{d-1} q'$. Therefore, the expectation value $\mathbb{E}(H_k)$ can be calculated as
    \begin{equation}
        \begin{aligned}
            \mathbb{E}(H_k) =& \sum_{d=1}^\infty \Pr(d) \mathbb{E}(H_k|d)\\
            =& \sum_{d=1}^{l} (1-q')^{d-1} q' [d + (1 - q_c) \mathbb{E}(H_k)] + \sum_{d=l+1}^\infty (1-q')^{d-1} q' [d + \mathbb{E}(H_k)] \\
            =& \sum_{d=1}^{\infty} (1-q')^{d-1} q' d + (1 - q_c) \mathbb{E}(H_k) \sum_{d=1}^{l} (1-q')^{d-1} q' + \mathbb{E}(H_k) \sum_{d=l+1}^\infty (1-q')^{d-1} q' \\
            =& \frac{1}{q'} + (1 - q_c) \mathbb{E}(H_k) [1 - (1 - q')^l] + \mathbb{E}(H_k) (1 - q')^l \\
            =& \frac{1}{q'} + \mathbb{E}(H_k) \{1 + q_c [(1 - q')^l - 1]\}.
        \end{aligned}
    \end{equation}
We can obtain the expected value as
\begin{equation}
    \begin{aligned}
        \mathbb{E}(H_k) &= \frac{1}{q' q_c [1 - (1 - q')^l]} = \frac{q'}{ (q_0^2 + q_1^2) [1 - (1 - q')^l]}.
    \end{aligned}
\end{equation}
In this way, the pairing efficiency of the pairing strategy in Box 7 can be given as
\begin{equation}
    r' = \frac{1}{\mathbb{E}(G_k)} = \bigg[ \frac{1}{q'} + \frac{q'}{q_0^2 + q_1^2} \frac{1}{1 - (1 - q')^l} \bigg]^{-1}.
\end{equation}

In the above analysis, we do not consider the non-asymptotic case, in which the numbers of different pairs should be balanced to obtain the optimal secret key rate. Below we give a brief analysis of how to balance different pairs. In step-ai, Alice could measure the auxiliary system with the following measurement operators
\begin{equation}
    \begin{aligned}
        M''_0 &= \sqrt{f_0} \ket{0}\bra{0} + \ket{1}\bra{1} + \sqrt{1-f_2}\ket{2}\bra{2},\\
        M''_1 &= \sqrt{1-f_0} \ket{0}\bra{0} + \sqrt{f_2}\ket{2}\bra{2}.
    \end{aligned}
\end{equation}
When $f_0,f_2 \rightarrow 1$, the operators $\{M''_0,M''_1\}$ are just the operators $\{M'_0,M'_1\}$ in Eq. (\ref{eqs4}). When $f_0 \rightarrow 1$ and $f_2 \rightarrow 0$, Alice and Bob's labels $\mathcal{L}_k^a,\mathcal{L}_k^b$ will only be 0. In this way, the pairing strategy in Box 7 just reduces to that in Box 2. By optimizing $f_0,f_2\in[0,1]$, the number of different pairs can be adjusted to obtain the optimal secret key rate in the non-asymptotic case. The security analysis for this pairing strategy is the same as above but with a little modification of the POVM with one more operator as $M_3 = \mathbb{I} - M_0 - M_1 - M_2$.

\section{Detailed Calculation}
\label{sup_det_cal}
We first show the detailed calculation of the first equation in Eq. (\ref{eqk}).

\begin{equation}
    \rho_1 = \frac{1}{2\pi} \int_0^{2\pi} \hat{P} \big(\ket{\varphi_{1,\theta+\delta,\theta}^\prime} \big) d\theta = \frac{1}{2} \sum_{s,t\in\mathbb{Z}_2} \ket{s\bar{s}} \bra{t\bar{t}} \otimes \rho_{s,t,\theta+\delta,\theta},
\end{equation}
where
\begin{equation}
    \begin{aligned}
        \rho_{s,t,\theta + \delta,\theta} &= \frac{1}{2\pi} \int_{0}^{2\pi} \ket{e^{i(\theta + \delta)}\sqrt{s\mu}} \bra{e^{i(\theta + \delta)}\sqrt{t\mu}} \otimes \ket{e^{i\theta}\sqrt{\bar{s}\mu}} \bra{e^{i\theta}\sqrt{\bar{t}\mu}} d\theta\\
        &= \frac{1}{2\pi} \int_{0}^{2\pi} \sum_{j,k,m,n \in \mathbb{N}} e^{-\mu} e^{i(j-k)(\theta + \delta) + i(m-n)\theta} \frac{\mu^{(j+k+m+n)/2} s^j t^k \bar{s}^m \bar{t}^n}{\sqrt{j!k!m!n!}} \ket{j,m}\bra{k,n} d\theta \\
        &= \sum_{j,k,m,n \in \mathbb{N}} e^{-\mu} e^{i(j-k)\delta} \frac{\mu^{(j+k+m+n)/2} s^j t^k \bar{s}^m \bar{t}^n}{\sqrt{j!k!m!n!}} \ket{j,m}\bra{k,n} \frac{1}{2\pi} \int_{0}^{2\pi} e^{i(j-k+m-n)\theta} d\theta \\
        &= \sum_{j+m = k+n} e^{-\mu} e^{i(j-k)\delta} \frac{\mu^{(j+k+m+n)/2} s^j t^k \bar{s}^m \bar{t}^n}{\sqrt{j!k!m!n!}} \ket{j,m}\bra{k,n} \\
        & \xlongequal[]{z \triangleq j+m = k+n} \sum_{z \in \mathbb{N}} e^{-\mu} e^{i(s-t)z\delta} \frac{\mu^{z}}{z!} \ket{sz,\bar{s}z}\bra{tz,\bar{t}z} \\
        & \xlongequal[]{n \triangleq z} \sum_{n \in \mathbb{N}} e^{i(s-t)n\delta} p_{\mu,n} \ket{sn,\bar{s}n}\bra{tn,\bar{t}n}.
    \end{aligned}
\end{equation}
Therefore, we could obtain the following formula
\begin{equation}
    \begin{aligned}
        \rho_1 &= \frac{1}{2} \sum_{s,t\in\mathbb{Z}_2} \Big[ \ket{s,\bar{s}}\bra{t,\bar{t}} \otimes \sum_{n \in \mathbb{N}} e^{i(s-t)n\delta} p_{\mu,n} \ket{sn,\bar{s}n}\bra{tn,\bar{t}n} \Big] \\
        &= \sum_{n \in \mathbb{N}} p_{\mu,n} \sum_{s,t\in\mathbb{Z}_2} \ket{s,\bar{s}}\bra{t,\bar{t}} \otimes e^{i(s-t)n\delta} \ket{sn,\bar{s}n}\bra{tn,\bar{t}n} \\
        &= \sum_{n \in \mathbb{N}} p_{\mu,n} \hat{P} \Big[ \frac{1}{\sqrt{2}} \big(\ket{0,1} \ket{0,n} + e^{in\delta} \ket{1,0} \ket{n,0}\big) \Big] \\
        &= \sum_{n \in \mathbb{N}} p_{\mu,n} \hat{P} (\ket{\varphi_{n,\delta}}).
    \end{aligned}
\end{equation}
This proves the first equation in Eq. (\ref{eqk}). The second equation in Eq. (\ref{eqk}) is
\begin{equation}
    \rho_2 = \frac{1}{2\pi} \int_0^{2\pi} \hat{P} \big(\ket{\varphi_{2,\theta+\delta,\theta}^\prime}\big) d\theta = \ket{11}\bra{11} \otimes \sigma_{\theta+\delta,\theta},
\end{equation}
where
\begin{equation}
    \begin{aligned}
        \sigma_{\theta+\delta,\theta} &= \frac{1}{2\pi} \int_{0}^{2\pi} \ket{e^{i(\theta+\delta)} \sqrt{\mu}} \bra{e^{i(\theta+\delta)} \sqrt{\mu}} \otimes \ket{e^{i\theta} \sqrt{\mu}} \bra{e^{i\theta} \sqrt{\mu}} d\theta \\
        &= \frac{1}{2\pi} \int_{0}^{2\pi} \sum_{j,k,m,n\in\mathbb{N}} e^{-2\mu} e^{i(j-k)(\theta+\delta) + i(m-n)\theta} \frac{\mu^{(j+k+m+n)/2}}{\sqrt{j!k!m!n!}} \ket{jm} \bra{kn} d\theta \\
        &= \sum_{j,k,m,n\in\mathbb{N}} e^{-2\mu} e^{i(j-k)\delta} \frac{\mu^{(j+k+m+n)/2}}{\sqrt{j!k!m!n!}} \ket{jm} \bra{kn} \frac{1}{2\pi} \int_{0}^{2\pi} e^{i(j-k+m-n)\theta} d\theta \\
        &= \sum_{j+m=k+n} e^{-2\mu} e^{i(j-k)\delta} \frac{\mu^{(j+k+m+n)/2}}{\sqrt{j!k!m!n!}} \ket{jm} \bra{kn} \\
        &\xlongequal[]{z \triangleq j+m = k+n} \sum_{z\in\mathbb{N}} \sum_{j,k=0}^{z} e^{-2\mu} e^{i(j-k)\delta} \frac{\mu^{z}}{\sqrt{j!k!(z-j)!(z-k)!}} \ket{j,z-j} \bra{k,z-k} \\
        &\xlongequal[]{n \triangleq z} \sum_{n\in\mathbb{N}} e^{-2\mu} \frac{(2\mu)^n}{n!} \sum_{j,k=0}^{n} e^{i(j-k)\delta} \frac{n!}{(2\mu)^n} \frac{\mu^{n}}{\sqrt{j!k!(n-j)!(n-k)!}} \ket{j,n-j} \bra{k,n-k} \\
        &= \sum_{n\in\mathbb{N}} p_{2\mu,n} \sum_{j,k=0}^{n} e^{i(j-k)\delta} \frac{1}{2^n} \sqrt{ C_n^j C_n^k }\ket{j,n-j} \bra{k,n-k} \\
        &= \sum_{n\in\mathbb{N}} p_{2\mu,n} \hat{P} \bigg[ \frac{1}{\sqrt{2^n}} \sum_{j=0}^{n} e^{i\delta} \sqrt{ C_n^j}\ket{j,n-j} \bigg] \\
        &= \sum_{n\in\mathbb{N}} p_{2\mu,n} \hat{P} (\ket{\omega_{n,\delta}}),
    \end{aligned}
\end{equation}
Hence, we obtain the following state
\begin{equation}
    \rho_2 = \sum_{n\in\mathbb{N}} p_{2\mu,n} \hat{P} \big(\ket{\phi_{n,\delta}}\big).
\end{equation}

\bibliographystyle{spphys}
\bibliography{mp}

\end{document}